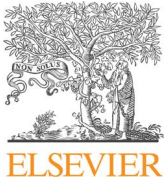

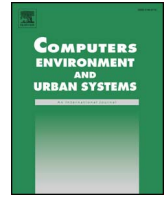

# Big enterprise registration data imputation: Supporting spatiotemporal analysis of industries in China

Fa Li[a,c], Zhipeng Gui[b,a,c,*], Huayi Wu[a,c], Jianya Gong[b,a,c], Yuan Wang[a,c], Siyu Tian[b], Jiawen Zhang[b]

[a] *State Key Laboratory of Information Engineering in Surveying, Mapping and Remote Sensing, Wuhan University, Wuhan, China*
[b] *School of Remote Sensing and Information Engineering, Wuhan University, Wuhan, China*
[c] *Collaborative Innovation Center of Geospatial Technology, Wuhan, China*

## ARTICLE INFO

## ABSTRACT

Big, fine-grained enterprise registration data that includes time and location information enables us to quantitatively analyze, visualize, and understand the patterns of industries at multiple scales across time and space. However, data quality issues like incompleteness and ambiguity, hinder such analysis and application. These issues become more challenging when the volume of data is immense and constantly growing. High Performance Computing (HPC) frameworks can tackle big data computational issues, but few studies have systematically investigated imputation methods for enterprise registration data in this type of computing environment. In this paper, we propose a big data imputation workflow based on Apache Spark as well as a bare-metal computing cluster, to impute enterprise registration data. We integrated external data sources, employed Natural Language Processing (NLP), and compared several machine-learning methods to address incompleteness and ambiguity problems found in enterprise registration data. Experimental results illustrate the feasibility, efficiency, and scalability of the proposed HPC-based imputation framework, which also provides a reference for other big georeferenced text data processing. Using these imputation results, we visualize and briefly discuss the spatiotemporal distribution of industries in China, demonstrating the potential applications of such data when quality issues are resolved.

## 1. Introduction

Big data with fine-grained street-level location and coordinates, as well as operating period and industrial category information can deepen and extend analysis of industrial spatial distributions, thereby promoting a deeper understanding of urban processes. The spatial distribution of various economic activities lies at the very heart of theories of urban spatial structure and is essential for rational urban and regional economic planning and policymaking (Li, Zhang, Chen, & Yu, 2015; Parr, 2014). However, due to the lack of complete, fine-grained micro-level enterprise or firm data, few studies have fully analyzed the spatial distribution of industries in China at multiple scales, from a temporally sensitive perspective, incorporating all kinds of enterprises (Watkins, 2014; Zhu & Chen, 2007). The local bureaus of Administration for Industry and Commerce (AIC) of China, are responsible for enterprise registration, supervision and administration, and protection of consumers' rights and interests (AIC, 2016). These regional bureaus record detailed operating information for each enterprise. Big enterprise registration data, collected from multiple regional bureaus of AIC of China, can enable and support spatial-temporal analysis of industries, if the data quality issues are resolved.

Incompleteness and address ambiguity are prominent quality problems of Chinese enterprise registration data. A typical registration record contains information of an individual enterprise, including enterprise name, address, registration date, industrial category, business scope, postcode, legal representative, and registered capital. Usually, these records are manually recorded and inputted into the system at local AIC offices. In this process, critical information is either overlooked or neglected, and therefore frequently missing from the database. For example, in our study, 43.64% of the data has no industrial category values. This information however, is imperative when executing a spatial distribution analysis of industrial categories and industries. Approximately 30% of the records only have a street-level address but do not include the province or city to which it belongs. This address ambiguity problem is defined as the missing Administrative Division (AD) information problem, seriously impeding effective geocoding (Roongpiboonsopit & Karimi, 2010). To obtain the complete and accurate industrial category values, and the multi-scale text address and coordinates for each enterprise, imputation is required when filling missing values and information (Luengo, García, & Herrera, 2012).

Imputation however, introduces troublesome computing challenges when data volume is big. Enterprise registration data is in a short text format and text-based data imputation involves Natural Language Processing (NLP) techniques, such as short text classification and matching. This process is computing intensive and may result in the Out Of Memory and Intolerable Calculation-Time problems on a stand-alone computer when data volume is big. High Performance Computing (HPC) frameworks are often used to handle the computational issues of big data (Yang, Huang, Li, Liu, & Hu, 2016). Previous research explored big text data processing based on HPC frameworks. However, few studies have systematically investigated HPC-based imputation for big georeferenced text data that involves short text classification, location imputation and geocoding. Moreover, the discussion and applications of such technologies in regional and social science is insufficient in literature.

To fill this gap in the research and solve the big data quality problems endemic to this enterprise registration data, we propose an imputation framework and develop parallel imputation methods based on cutting-edge HPC technologies, to make this data more applicable. An effective solution to these kinds of data quality problems is relevant in many other domains where the use of big data is impeded by incompleteness and the ambiguity issues, especially for big georeferenced text data classification and location geocoding. We compare several widely used text classification methods employing NLP based on Apache Spark to fill missing industrial category values, in terms of accuracy, execution time, memory consumption, and scalability. We also introduce a location imputation method to fill the missing location information and obtain coordinates of each enterprise. Using these imputation results, we briefly analyze the spatiotemporal distribution of all industries in China at multiple spatial scales to illustrate potential applications of this data for analysis of urban spatial structures, urban agglomerations, industrial aggregations, and socioeconomic activities.

This article is organized as follows. Section 2 reviews relevant research. Section 3 introduces the data and HPC-based imputation framework. Section 4 describes industrial category and location imputation. Section 5 details the data imputation experiments and briefly analyzes the potential applications of generated data. Section 6 concludes this article and discusses future research.

## 2. Related work

### 2.1. Industrial spatial distribution analysis

Analysis of industrial spatial distribution has been highlighted in economic geography, urban spatial structure, and regional policy studies. Substantial studies on economics have analyzed the geographical concentration of industries, and the effects of agglomeration economies (Combes, Duranton, Gobillon, & Roux, 2008; Puga, 2010); by analyzing the industrial spatial distribution, many scholars have tried to explain the urban spatial structures (Giuliano & Small, 1991; Liu & Wang, 2016), improve the land use efficiency (Huang, He, & Zhu, 2017), as well as reveal the impact of regional policy on economic activities (Li et al., 2015). In these studies, enterprise or firm data is widely used, including aggregated data, and micro enterprise data. As distinct from aggregated data, micro enterprise data allows users to analyze information at varying spatial levels or partitions, and provides much more fine-grained individual information, offering the potential for theoretical innovation in economic geography and regional studies that are invisible in aggregated data sets (Domenech, Lazzeretti, Molina, & Ruiz, 2011; Lennert, 2011).

The use of micro enterprise data is a promising path in studies of industrial spatial distributions. In economic geography, based on micro enterprise data, distance based spatial agglomeration or clustering of enterprises (Duranton & Overman, 2005; Marcon & Puech, 2010), enterprise heterogeneity (Bernard, Jensen, Redding, & Schott, 2011), and continuous spacial modeling of enterprises (Arbia, 2010) are an active area of research. Domenech et al. (2011) conclude that the use of micro-data allows for much richer and detailed results than classical approaches with aggregated data. Using micro data, studies showed how economic activities were shaped by government and market forces (Li et al., 2015), and urban planning suggestions were proposed to optimize the urban spatial structure for achieving greater efficiency (Zhu & Chen, 2007). Many insightful case studies on urban spatial structure and urban expansion have been completed at the inter-city, regional (Liu, Derudder, & Wu, 2016), city and intra-city scales (Gao, Huang, He, Sun, & Zhang, 2016). Further studies are required to examine subcenters within urban districts (Liu & Wang, 2016). In addition, spatiotemporal distribution analysis at multiple scales demands comparison of different regions and industries, as spatial distributions of different industries vary over time and across space (Kneebone, 2010). Industrial spatial distribution analysis therefore, needs chronological, micro-level enterprise data with reliable, accurate, and complete industrial category and location information.

Currently, there have been insufficient studies of the spatial distribution of industries from a multi-scale and temporally sensitive perspective incorporating all kinds of enterprises. It is probably due to the lack of complete data for enterprises containing industrial operating periods, industrial categories and precise location information in the form of geographical coordinates. Big enterprise registration data collected from multiple regional AIC bureaus of China will enable and support such analysis; however, effective use of this data is impeded by data quality problems. To fill this gap in the research, we propose a HPC-based imputation framework to solve the big data quality problem endemic to enterprise registration data, making this data more readily applicable to researchers, planners and decision-makers.

### 2.2. Data imputation in big data era

Imputation has been widely used to fill the missing values in various data types. Datasets, including numerical data and text data, are prone to missing value problems. For numerical data imputation, mathematical methods can be directly used to provide the estimated values for incomplete data by analyzing the relation between multiple data fields or the relationship between different records in the same data field (Luengo et al., 2012; Sim, Kwon, & Lee, 2016). Different from numerical data imputation, text data imputation can harness NLP for semantic analysis. For text data imputation, to label a text with predefined categories, text classification is required; to extract unambiguous location information and even accurate coordinates from georeferenced text data, location estimation and geocoding are often needed (Chen, David, & Yang, 2013; Lennert, 2011). For example, there is a need to classify, estimate and geocode text location for social media data (Barapatre, Meena, & Ibrahim, 2016; Ghahremanlou, Sherchan, & Thom, 2015; Krumm & Horvitz, 2015). Classification, location estimation and geocoding are quite important to georeferenced text data processing.

Short texts are more intractable to be processed than normal document. Short texts are much shorter, nosier, and sparser. For example, a tweet has at most 140 characters (Sun, 2012) and it does not provide sufficient word occurrence, thus impeding traditional text representation methods for classification, such as "bag of words" model (Sriram, Fuhry, Demir, Ferhatosmanoglu, & Demirbas, 2010). For short text category imputation, most existing approaches try to enrich the representation of a short text using additional semantics derived from external sources such as Wikipedia and WordNet (Hu, Sun, Zhang, & Chua, 2009). These methods are limited by the completeness of external corpus, and the lack of semantic-consistency between external corpus and the classified short texts (Zhang & Zhong, 2016), especially in domain-specific studies (Wu, Morstatter, & Liu, 2016). The accuracy of short text classification using few words extracted directly from short texts may be limited in some cases. However, this method is simple and applicable when the existing semantic-consistent external sources are insufficient. Few researches have systematically explored short text classification without external sources (Sun, 2012). For text location imputation and geocoding, a complete gazetteer, postal data, street network data, parcel boundary data, or point data is required (Zandbergen, 2008; Goodchild & Hill, 2008; Melo & Martins, 2017). However, in China, there is no complete, public gazetteer, or users may not have the access to the authoritative data sources (Zhu, 2013). Research on matching and imputation of manually-recorded unstructured incomplete text address based on web-crawled and fused structured data, is insufficient (Curriero, Martin, Boscoe, & Klassen, 2010; Mccurley, 2001). Short text classification involves high dimensional vector computing (Song, Ye, Du, Huang, & Bie, 2014), while location imputation that integrates external sources increases the data volume (Huang, Cao, & Wang, 2014). These issues introduce troublesome computing challenges when data volume is big.

HPC technologies were developed for tackling computationally complex problems and could enable big text data imputation. The open-source projects, including Apache Hadoop, Spark, and Flink, are the cutting-edge HPC frameworks for handling big data (Yang, Yu, Hu, Jiang, & Li, 2016). Apache Spark is a clustering computing framework that is more efficiency than Hadoop due to in-memory data caching and iterative processing strategies (Arias, Gamez, & Puerta, 2017). It powers a stack of libraries including SQL, graph computing, stream data processing, and machine learning. Meanwhile, Spark provides a powerful software ecosystem of libraries and tools with wide and active user communities. Comparing with Grid computing frameworks, such as Globus and Condor-G, it is easier to write Spark applications since the low-level communication, coordination, and supervision details are hidden by the framework (Apache Spark, 2016). Big text data classification involves high dimensional feature vectors. Thousands of dimensions and even higher dimensional features are often encountered in unstructured text data processing, often causing the Out Of Memory and Intolerable Calculation-Time problems (Weinberger, Dasgupta, Langford, Smola, & Attenberg, 2009; Zhao, 2011). Previous researchers explored the feasibility, efficiency (Choi, Jin, & Chung, 2016), and scalability (Semberecki & Maciejewski, 2016) of big text data classification based on HPC frameworks (Wei, Shi, & Su-Qin, 2012). However, few studies have compared and analyzed short text classification methods with thousands of features and multiple classes in Spark-based HPC environments in terms of accuracy, execution time, memory consumption, and scalability.

In this paper, we propose an imputation framework based on HPC technologies for efficient big textual enterprise registration data processing, that also provides a reference for other big georeferenced text data processing. To find a suitable method to fill the missing industrial category, several short text classification-based imputation methods employing NLP were analyzed and compared based on Apache Spark. To obtain a complete and accurate multi-scale text address and coordinates for each enterprise, a location imputation method is proposed that integrates external postcode data source, and employs NLP and public web Application Programming Interfaces (APIs). The data analysis shows that the imputation result can support industrial spatial distribution analysis from a multi-scale, chronological, and comprehensive industrial viewpoint.

## 3. Dataset and HPC-based imputation framework

### 3.1. Dataset description and imputation target

To obtain the industrial category, and the multi-scale address for each enterprise, missing value issues must be resolved. In our study, there were 16,676,304 enterprise registration records recorded from 1960 to 2015 and collected from multiple regional AIC offices across

**Table 1**
Format and missing percentage of enterprise registration data.

| Fields name | Enterprise name | Industrial category | Address | Postcode | Data source |
|---|---|---|---|---|---|
| Type | String | Enumeration (16 classes) | String | String | String |
| Missing (%) | 0.03 | 43.64 | 0.19 | 25.75 | 31.06 |
| Example | 武汉***物业管理有限公司 (Wuhan *** Estate Management Ltd.) | 房地产业 (Real estate) | 南京路16号 (16, Nanjing Road) | 430014 | 2004年注册_湖北 (Registered in Hubei, 2004) |

*Note:* for privacy issue, the sensitive keywords in the enterprise name that may help to identify the individual enterprise are replaced with ***.

Mainland China. These records were primitively stored in multiple excel files, and then merged into a relational database. Table 1 lists the names, types, percentage of missing values, and an exemplary case of the data fields that relate to industrial category and address information of an individual enterprise.

In Table 1, fields containing industrial category information, including enterprise name and industrial category fields, are used for industrial category imputation. According to the National Economic Industry Classification Standard (1994) of China, all industries are categorized into 16 primary categories. The 16 industrial categories and their symbols are listed in Table A in the Appendix A. However, based on our statistic, we found that 43.64% of records have no category values in the industrial category field. According to stipulations of the law on the Administration of Enterprise Name Registration of China (AIC, 2016), the name of an enterprise must contain characteristic industry information; the enterprise name therefore, can be used to fill the missing industrial category values.

The enterprise name, address, postcode, and data source fields, contain location information and can be used for location imputation. In the address field, although only 0.19% of the records have no address, there are about 30% of the records only have street information and thus cannot identify the city to which it belongs. For example, many cities in China all have a street named "Nanjing Road" but from the address field alone, the city it belongs to is uncertain, as shown in Table 1. This address ambiguity problem stems from the absence of AD information in the address field, including province, city, and county information. A postcode can help to identify the AD as long as the associations between the AD and postcode can be obtained. However, 25.75% of the raw data has no postcode. Moreover, the raw data has no coordinates for map visualization and regional spatial analysis. Location imputation therefore, uses location related fields to obtain an accurate and multi-scale location for each enterprise, including province, city, county, street, and coordinates.

In conclusion, two imputation tasks are involved, industrial category imputation, and location imputation that allow us to fill the missing industrial category values and obtain the multi-scale address of each enterprise.

### 3.2. Imputation workflow and computational framework

The proposed imputation workflow of industrial category imputation and location imputation, and the adopted computational frameworks are illustrated in Fig. 1. Industrial category imputation is a short text classification problem (Song et al., 2014) and consists of two-step, input vector construction and classification. The Apache Spark framework has advantages in performance with many machine learning methods in text classification, and therefore is employed to achieve industrial category imputation in parallel (Choi et al., 2016). Location imputation obtains an accurate multi-scale address for each enterprise, involving three steps, postcode imputation, AD imputation, and geocoding. For location imputation, the records are mutually independent; data parallelism strategy therefore, can be adopted to efficiently impute location with a bare-metal cluster. The bare-metal computing cluster is composed of multiple stand-alone virtual machines (VMs) without employing any advanced parallel computing framework. In this process, each stand-alone VM runs imputation program and processes a data piece independently without intermedia data exchange between different computing nodes. Details of industrial category imputation and location imputation will be discussed further in Section 4.

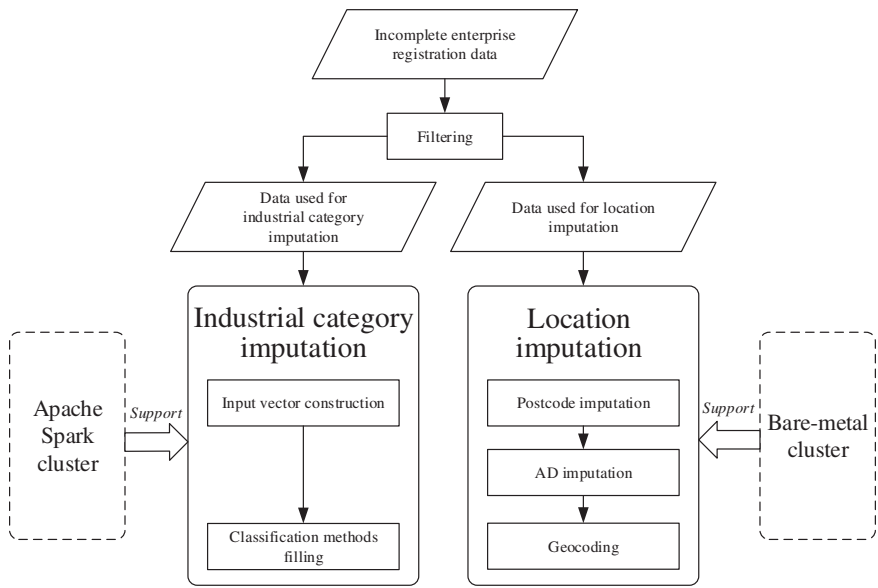

**Fig. 1.** Workflow and computational framework for imputation of incomplete enterprise registration data.

## 4. Industrial category imputation and location imputation

### 4.1. Industrial category imputation

Fig. 2 shows the sub-steps of input vector construction and classification in industrial category imputation workflow. The input vector construction is to transform each textual enterprise name into a digital vector for classification. The classification is to fill the missing industrial category values by training the correspondence between an input vector and industrial category using records with industrial category values.

Complex and high-dimensional computing with big volume data is involved in the process of industrial category imputation. On the left of Fig. 2, feature words are the words that refer to industrial characteristics information (Sun, 2012). Since there are more than ten million enterprise records in the dataset, the number of feature words extracted from the enterprise names is extremely large. High-dimensional feature words makes the process of classification both data-intensive and compute-intensive (Weinberger et al., 2009). Hence, the Apache Spark-based framework is essential to support this process (GreenC bot, 2016).

### 4.1.1. Input vector construction

The detailed workflow and an illustrative instance of input vector construction are illustrated in Fig. 3. In the figure, word segmentation transforms the Chinese string into words to achieve Part-Of-Speech (POS) tagging. We used the Ansj library (NLPChina, 2014) to achieve word segmentation; it is an efficient Java-based cross-platform word segmentation library, suitable for use in Apache Spark. In the case shown in Fig. 3, Wuhan is marked as *ns* representing an address noun, while the feature words for estate, and management, are marked as nouns (*n*) or verbs (*v*). By analyzing all of the samples, we selected 29,377 words whose POS were tagged as *n* (noun), *v* (verb), or *vn* (gerund) from all segmented words. We treat these selected words as the feature words, and use them for feature vector generation.

The feature hash method is used to generate the feature vector for classification. Each enterprise name is represented by textual feature words while the input for classification is a vector (Qwertyus, 2015). Since there are 29,377 feature words, the dimension of the constructed vector will be as high as 29,377 when constructed directly. Therefore, dimensionality reduction is applied to reduce the computational complexity of the classification method (García, Luengo, & Herrera, 2016). Feature hash is a method that transforms text words into vectors according to the hash code of each word, and efficiently executes dimension reduction (Weinberger et al., 2009). Using this method, we constructed each enterprise name into a vector with 15,000 dimensions, shown in Fig. 3. Since the industrial information contained in each enterprise name is limited and short, industrial category imputation becomes a short text classification problem (Wang et al., 2009); the classification methods therefore need to be carefully selected.

**Fig. 2.** Industrial category imputation workflow.

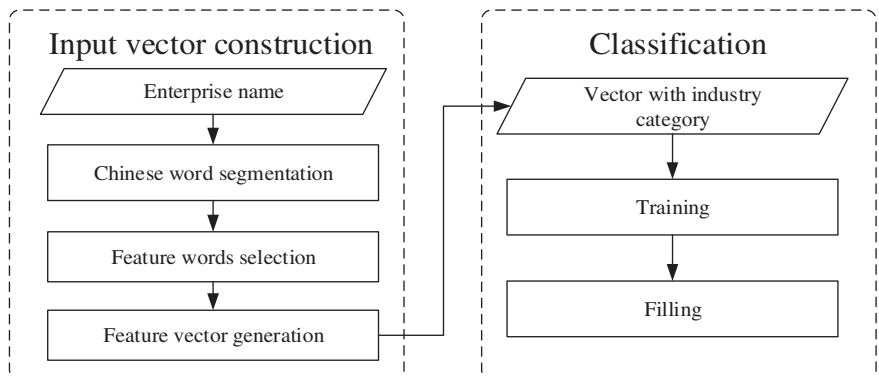

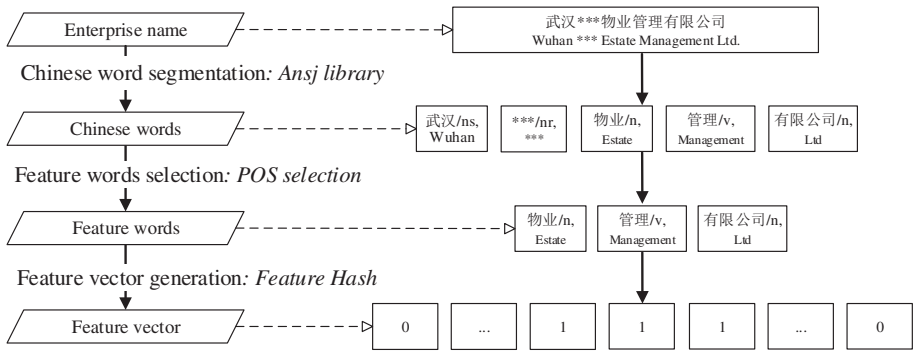

**Fig. 3.** Workflow and an illustrative instance of input vector construction.

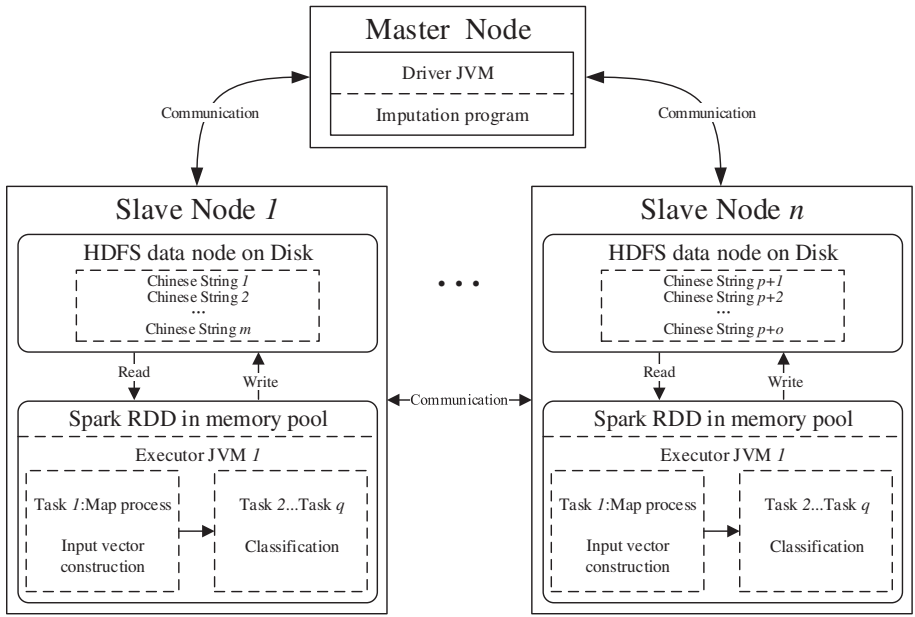

**Fig. 4.** Spark-based computing framework for industrial category imputation.

#### 4.1.2. Selection of classification methods

Classification methods must be carefully selected for the short textual category imputation problems with big volume. High-dimension and data-intensive computing are involved in the imputation, and challenge the applicability and efficiency of classification methods (Xu et al., 2015). The challenges are as follows:

1) *Sparse and high-dimensional classification*: the input vector for short text classification is high-dimension and sparse. Since each distinct word defines an independent feature, a feature vector is highly dimensional when a training dataset is large (Weinberger et al., 2009). These vectors are also extremely sparse according to Zipf's law (Qwertyus, 2015). High and sparse dimensionality leads to complex computing problems and lower accuracy for methods based on similarity of term frequency, such as KNN (Song et al., 2014). Hence, a classification method suitable for handling highly dimensional and sparse input vectors is needed to achieve higher classification accuracy.

2) *Probability-based or nonlinear classification*: different industrial categories may have the same feature words in the enterprise name. For example, "Commerce Co., Ltd" is a common word for an enterprise name in wholesale, retail trade and catering, as well as social service industrial categories. This is caused by the fuzziness and generality of natural language, and classification therefore, is probability-based or nonlinear (Song et al., 2014).

3) *Influence of noise features*: noise features are present in the input attributes because of the error in feature words selection (García et al., 2016). Reducing the impact of noise features is reasonable for better classification.

4) *Computability problems*: high dimensional and nonlinear (or probability-based) classification may cause computational problems. As time and space complexity is related to the number of features, some nonlinear or probability-based classification methods may be prohibitive for high dimensional data (Klopotek, 2002). These issues become more challenging for big data, as the complexity is also linearly, quadratically, or even exponentially related to the data size (Johnson & Papadimitriou, 1994). Classification methods therefore, need to be compared for time and space consuming, and the computational framework needs to be scalable when the data size grows.

To address these challenges and select the most applicable classification method, we compare and evaluate different short text classification methods, in terms of accuracy, execution time, memory consumption, and scalability in Section 5.1.1.

#### 4.1.3. Apache Spark-based imputation

The Apache Spark-based computing framework applied to the industrial category imputation problem is illustrated in Fig. 4. In this framework, the master node is used for task scheduling, sending the imputation program to slave nodes and driving them to execute the paralleled imputation work. As shown in Fig. 4, the big data are divided into many small pieces and distributed to HDFS dispersed on different slave nodes. HDFS stores the big data used for industrial category imputation and provides high throughput data access for Apache Spark. Each piece of data is loaded into a memory pool as Apache Spark Resilient Distributed Dataset (RDD) objects, to efficiently complete imputation. In the memory pool of a slave node, the map-reduce task is executed by the executor to achieve the vector construction and classification. The executor is an Apache Spark process and task is a thread running in memory. During the whole process, communication among the master node and the slave nodes is executed; the system records the execution status, and sends the intermediate data results along to the nodes, for further processing.

Following industrial category imputation, we fill the missing industrial category values. The remaining information missing problems are handled by the location imputation procedure.

### 4.2. Location imputation

The goal of location imputation is to obtain the precise address of each enterprise including province, city, county, street, and geographical coordinates. The location imputation workflow is illustrated in Fig. 5. In the process, an external postcode database is constructed through web crawler to support the postcode imputation and AD imputation. Postcode imputation fills in the missing postcode values based on the location fields and external postcode database. AD imputation fills in the missing AD information to obtain accurate multi-scale text addresses. Geocoding obtains the geographical coordinates of each enterprise. The details of each step are further discussed in the following sub-sections.

#### 4.2.1. Postcode database construction and postcode imputation

A postcode database was constructed to support postcode and AD imputation by integrating three external postcode data sources, including Postcode Library, Baidu Postcode, and China Post (see Fig. 5, and Table B in Appendix A). Postcode Library provides postcodes, and detailed but unstructured text addresses; Baidu Postcode provides structured text addresses, but requires complete postcodes or addresses as input; China Post provides authoritative and relatively complete postal data (Province-level, city-level, county-level addresses, street, block, and township are completely covered), but it has restrictive query limitations to impede web crawling. To overcome these limitations, postcodes crawled on Postcode Library are used to query Baidu Postcode; while China Post is used to validate the postcodes. By using this integration strategy, a relatively complete database with more than 1.7 million postcodes and their corresponding AD records was constructed. Each postcode and its AD were recorded as shown in Table 2.

By matching postcodes and text addresses of enterprises to the postcode database, the validity and coverage of the constructed postcode database are evaluated. The postcode database covers all postcodes with correct postcode format in enterprise registration database. 96.34% of enterprise registration records with complete AD and postcodes can correctly matched the four-level AD (i.e., province, city, county and street) in postcode database by using the address tree-based Vector Space Model (VSM) (Kang, Du, & Wang, 2015). The high match rate demonstrates the relative completeness and reliability of the constructed postcode database.

**Table 2**
The correspondence between postcode and AD in postcode database.

| Record ID | Province | City | County | Street | Postcode |
|---|---|---|---|---|---|
| 1 | 湖北省 Hubei | 武汉市 Wuhan | 江岸区 Jiangan District | 南京路 Nanjing Road | 430014 |

Using this external postcode database, missing postcodes are filled. In the case illustrated in Fig. 6, the postcode of an enterprise record is missing but the location fields including data source, address, and institution name fields have implicit AD information. In this example, address nouns marked as *ns* are extracted through word segmentation from these fields and the part-of-speech selection process. By using these extracted address nouns, SQL fuzzy matching efficiently searches for corresponding records in the postcode database. If no record or multiple records are matched through SQL, the record with the maximum matching degree using the address tree-based VSM is selected as the matched record. If multiple records achieve the maximum matching degree (there is only a 0.3% chance in a search), we use these address nouns to match the registration records with postcodes in the enterprise database, and calculate the appearance probability for each postcode using Formula 1. The postcode with maximum probability is selected to fill the missing postcode. Using postcodes, AD imputation and geocoding can be achieved.

$$P(i) = \frac{N_i}{\sum_{i=1}^{n} N_i} \quad 1 \leq i \leq n \tag{1}$$

where $P(i)$ is the appearance probability of each matched postcode; $n$ is the number of matched postcodes; $N$ is the number of the enterprises with the $i$-th postcode and the matched addresses nouns in the enterprise registration database.

#### 4.2.2. Administrative division (AD) imputation and geocoding

The AD imputation and geocoding workflow is illustrated in Fig. 7. First, the postcode of each enterprise is selected from the enterprise database. Second, AD of each enterprise is obtained by searching its postcode in the postcode database. Then, the unambiguous address of each enterprise is obtained by combining the AD and ambiguous street-level address. Finally, we assign geographical coordinates to each unambiguous text address by invoking a public geocoding API.

The computing framework for AD imputation and geocoding is illustrated in Fig. 8. To obtain the coordinates, the online geocoding API is invoked using the API key obtained from the API providers. Each key permits a limited number of requests per day. For example, Baidu Geocoding API V2.0 has a usage limitation of 6000 times per day (Baidu, 2016). Due to this restriction and the huge number of text addresses that need geocoding, multiple developer accounts with several API keys are necessary. As shown in Fig. 8, to achieve efficient AD imputation and geocoding of millions of text addresses, multiple VMs are employed to process simultaneously this big text data. The dataset is evenly divided into many pieces and distributed to different VMs where a unique geocoding API key is assigned on each VM.

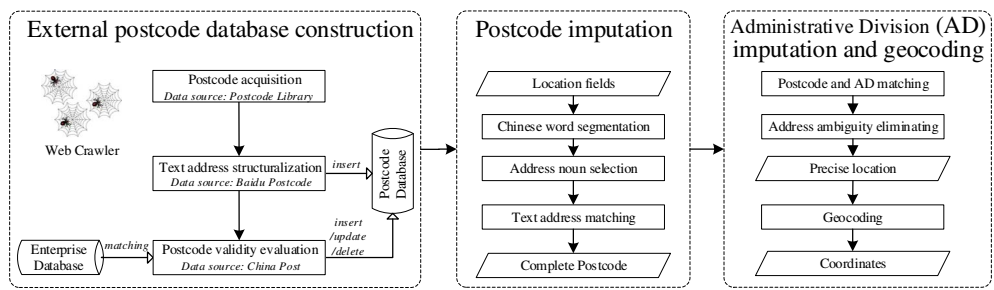

**Fig. 5.** Location imputation workflow.

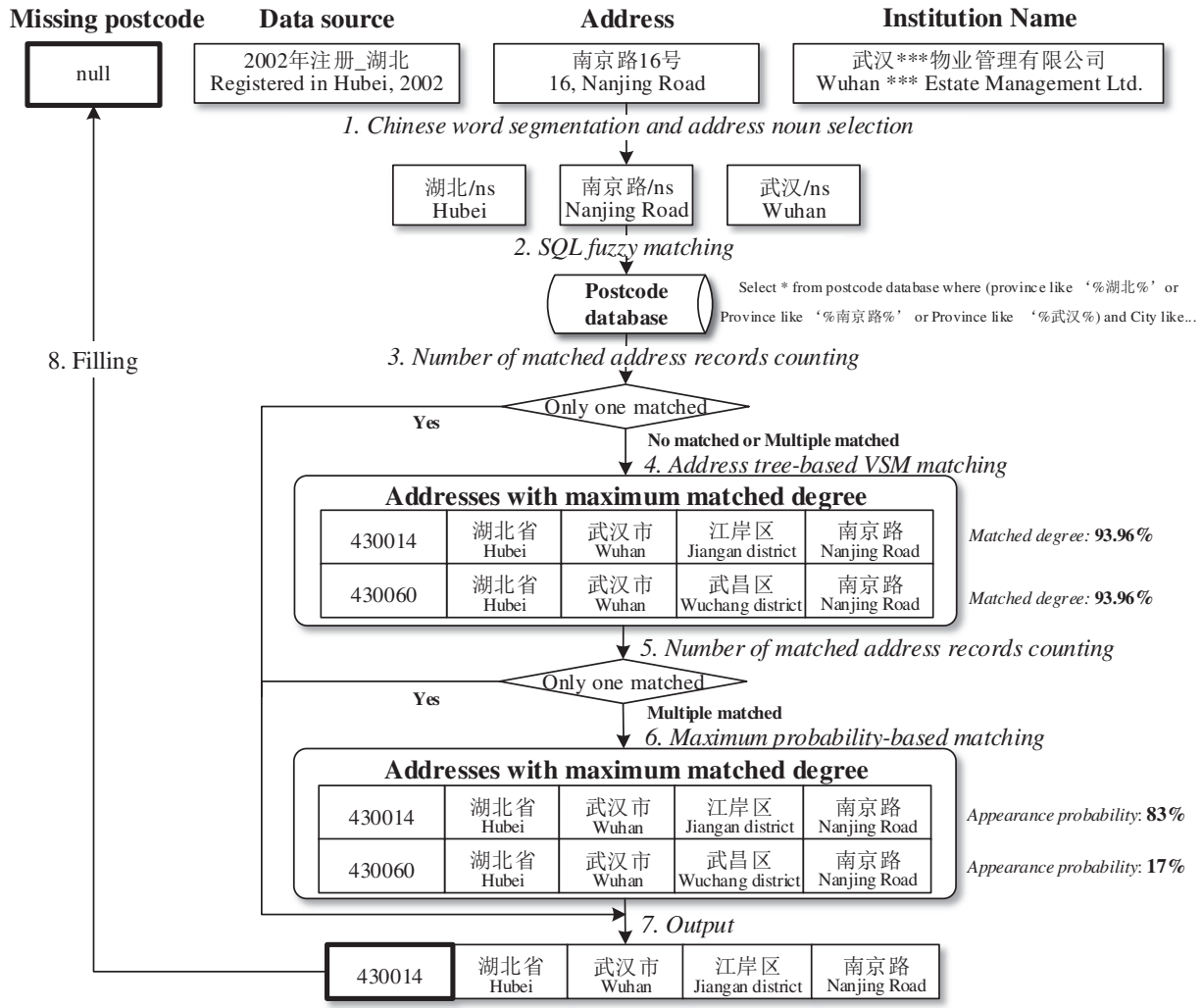

Fig. 6. Workflow and an illustrative instance of postcode imputation.

In Section 4, we detailed the workflow and computing framework for industrial category imputation and location imputation. Experiments to validate the workflow and framework are discussed in Section 5.

Fig. 7. Workflow of AD imputation and geocoding.

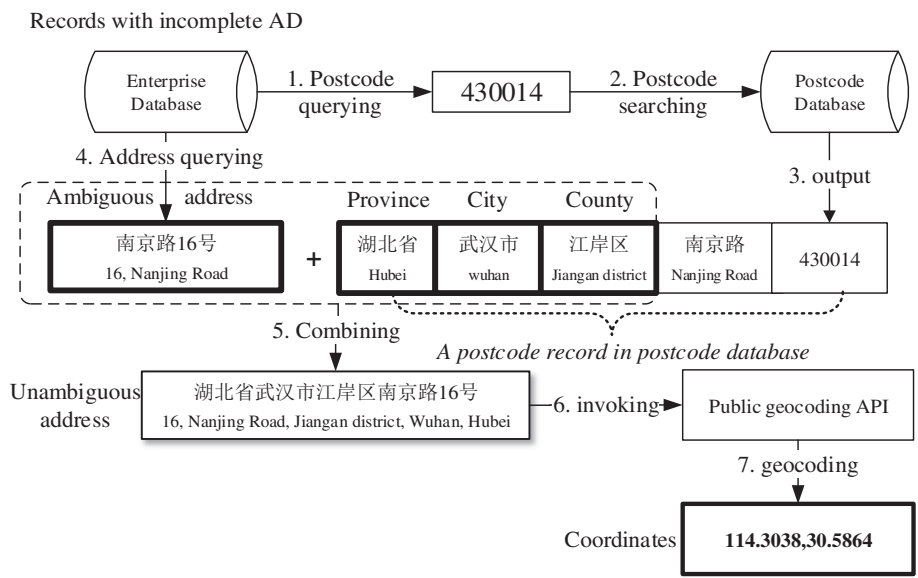

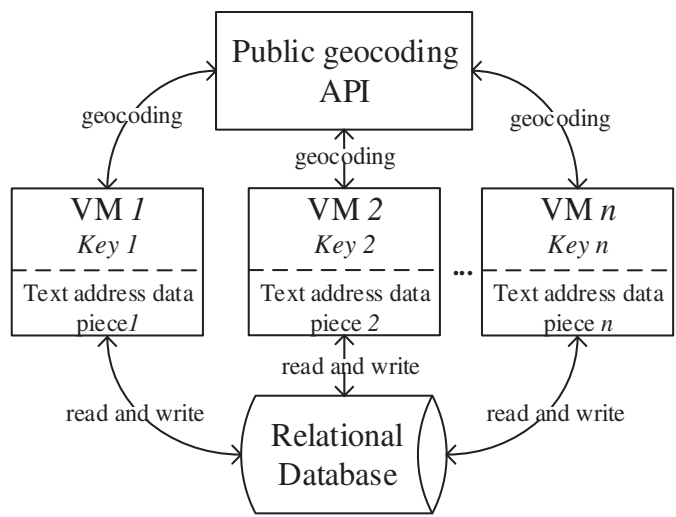

**Fig. 8.** Computing framework of AD imputation and geocoding.

## 5. Imputation experiments and potential application analysis

Imputation experiments were conducted to verify the feasibility and effectiveness of the proposed workflow and computing framework for industrial category imputation and location imputation. After presenting the imputation experiments, the spatiotemporal distribution of industries in China will be briefly analyzed, revealing the potential applications of such a data when quality issues are resolved.

### 5.1. Imputation experiments

We conducted two groups of experiments for industrial category imputation and two experiments for location imputation respectively. In industrial category imputation experiments, the first group of experiments compared different classification methods on Apache Spark cluster for accuracy, execution time, and memory consumption. The second group revealed the scalability of imputation framework through comparing the execution time and speed-up ratio on different number of slave nodes in a Spark cluster. After these two groups of experiments, the results of industrial category imputation were analyzed. In location imputation experiments, the accuracy of postcode imputation and AD imputation were calculated and deviations of geocoding results were analyzed.

The configuration of the imputation framework as used in the experiments was as follows. The Apache Spark cluster used in classification method comparison experiments was composed of four VMs, a single master node with three slave nodes. The configuration of each VM is 32 GB memory. While in scalability experiments, the number of slave nodes varied from one to sixteen. The configuration of each VM in scalability experiment and location imputation experiments with bare-metal cluster was 16 GB memory. Each VM had a 4x2GHz CPU, and Centos 6.5(64-bit) OS.

#### 5.1.1. Industrial category imputation experiments

To fill the missing industrial category values using the most suitable classification method, five classification methods contained in the Apache Spark MLlib library, were selected and compared. These methods are widely used in text classification, including Naïve Bayes, Random Forest, Decision Tree, Support Vector Machine (SVM), and Logistic Regression (Luengo et al., 2012; Song et al., 2014).

To estimate the imputation accuracy, ten-fold cross-validation was used on each data set (Kohavi, 1995). 9,398,927 records with industrial categories were selected to create a dataset without missing categories. To validate the efficiency of Apache Spark when handling different data size, the dataset was randomly divided into sub-datasets with different size and the proportion was used to mark each divided sub-dataset. For example, 0.1 means one-tenth of the records were selected. The results of comparative experiments are discussed as follows.

(1) Comparison of accuracy, execution time and memory consumption for five classification methods

Accuracy, execution time, and memory consumption results for Naïve Bayes, Random Forest, Decision Tree, SVM, and Logistic Regression methods are shown in Fig. 9(a–c). Since the settings of model parameters for each classification method are beyond the scope of this paper, we show the selected parameters in Table C in Appendix A. These parameters are set to achieve higher accuracy and consume less execution time in our experiments. As illustrated in Fig. 9(a–c), the accuracy of Naïve Bayes and Logistic Regression is much higher than other methods. Here we use a linear SVM for comparison. The low accuracy of linear SVM suggests that the imputation problem is non-linear or probability-based as mentioned in Section 4.1.2. Non-linear SVM may provide higher accuracy, but is much more time and memory consuming, as it uses a kernel function to solve none-linear problem by transforming this high-dimensional problem into a higher-dimensional problem (Jebara, 2004). The execution time and memory consumed in Naïve Bayes is much less than the other four methods. Experiments show that the memory usage in Random Forest and Decision Tree reaches about 53 GB. SVM and Logistic Regression reach about 116 MB and 178 MB when the data size is 0.1 (about 0.94 million records), while the memory usage of Naïve Bayes is only 15.8 MB. However, the accuracy of Logistic Regression is a bit higher (about 3%) than Naïve Bayes.

Experiments and analysis demonstrate that Naïve Bayes and Logistic Regression are more suitable for short textual classification with highly dimensional sparse vector input. They can solve imputation problems at much higher accuracy, with less execution time, than the other three methods. To select the better classification methods, the scalability experiments of Naïve Bayes and Logistic Regression were conducted using the entire dataset.

(2) Scalability experiments

The results of scalability experiments for Naïve Bayes and Logistic Regression using the entire dataset, are shown in Fig. 10. The Speed-up ratio is the execution time of one slave node divided by the execution time of clustered VMs. In Fig. 10, the execution time decreases and speed-up ratio increases as the number of slave nodes increases from one to fourteen, revealing the scalability of Spark-based imputation framework. When the number of slave nodes is sixteen, the execution time increases and speed-up ratio decreases. The efficiency decline is caused by the increasing schedule cost. Overall, the Spark-based imputation framework is scalable, but need to consider the tradeoff between the computing cost and communication or scheduling cost.

To select a better method, we can make a tradeoff between accuracy, time and resource consumption. In Fig. 10, it costs 39 s to achieve about 75% average accuracy using Naïve Bayes and 14 slave nodes. Naïve Bayes is qualified for near real-time big text data classification based on Apache-Spark. Compared to Naïve Bayes, the average accuracy of Logistic Regression is about 3% higher than that of Naïve Bayes; nevertheless, the execution time of Logistic Regression was about 30 to 40 times longer than Naïve Bayes, and memory usage was about 10 times larger than Naïve Bayes. For enterprise registration data imputation, it is one-time running task; the time and memory consumption of Logistic Regression is acceptable. To achieve higher accuracy, we selected the Logistic Regression method. The pseudocode for the Logistic Regression imputation method in the Apache Spark environment is shown in Table D in Appendix A.

(3) Accuracy analysis of imputation results

The Logistic Regression method was selected to complete the industrial

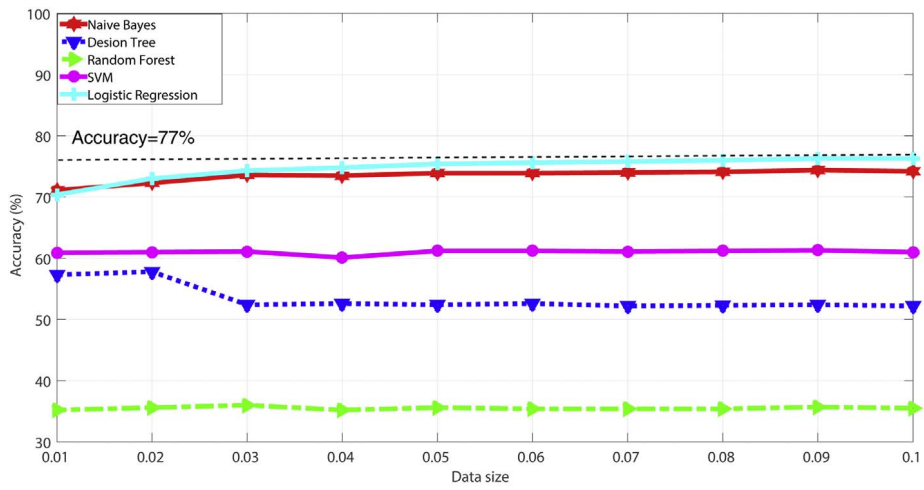

(a) accuracy comprison

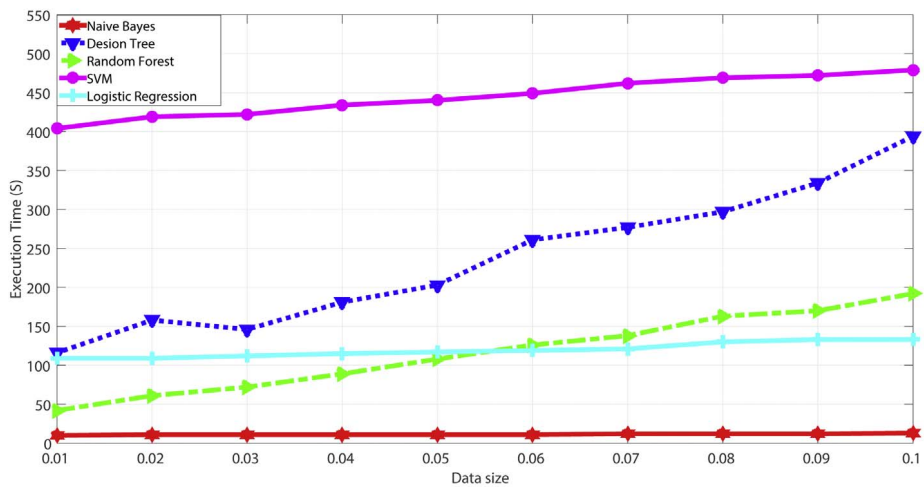

(b) execution time comparison

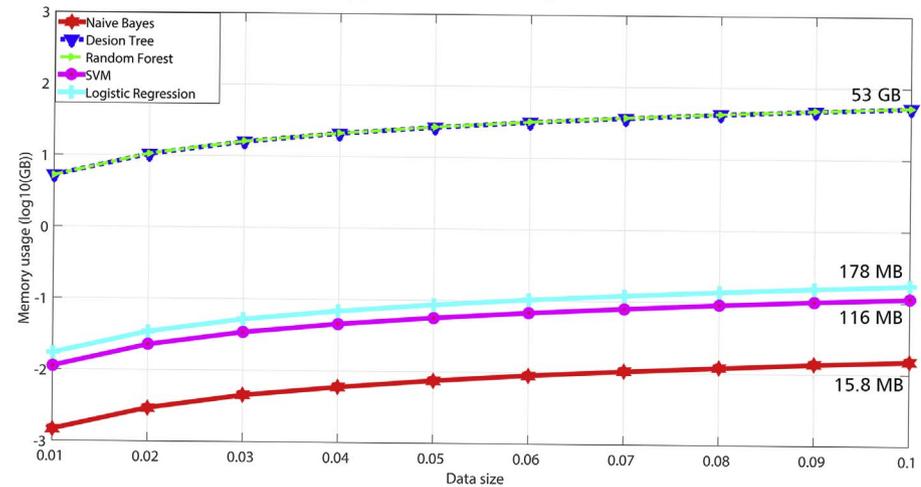

(c) memory usage comparison

category imputation. Based on Apache Spark, Logistic Regression was firstly validated using 9,398,927 records with category values, and then was used to fill 7,277,377 records with no industrial category values. The classification accuracy and proportion of each industrial category is shown in Fig. 11. This figure shows that some industrial categories achieve more than 90% classification accuracy. Nevertheless, some industrial categories have lower classification accuracy. Even though the industrial categories with lower accuracy occupy only a small proportion of the total records, it is necessary to reconsider and explain the reason why accuracy is lower. Industrial category imputation is a short text classification problem and the industrial information contained in the institution name is limited. For the lower accuracy categories, it is necessary to integrate other data sources containing industrial category information to improve the classification accuracy. Overall, it is feasible to achieve about 77.4% average accuracy for industrial category imputation using only institution name information.

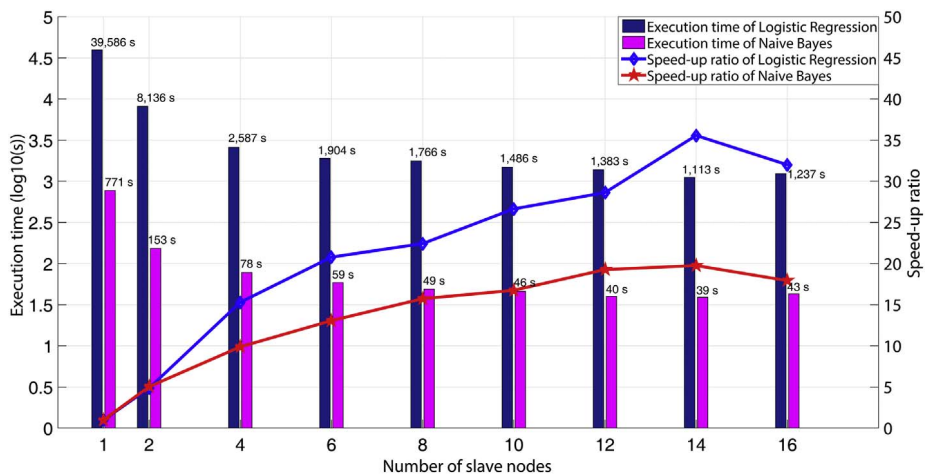

Fig. 10. Scalability experiments of Naïve Bayes and Logistic Regression using Apache Spark.

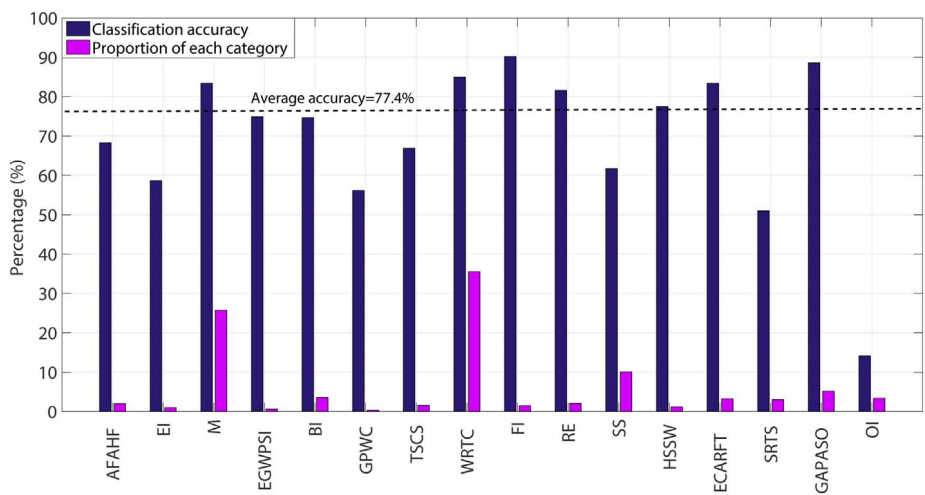

Fig. 11. Industry category imputation result (The full name of each industrial category abbreviation can be seen in Table A in Appendix A).

### 5.1.2. Location imputation experiments

In location imputation experiments, the accuracy of postcode imputation and AD imputation were calculated and deviations of geocoding results were analyzed. These experiments demonstrate the feasibility of postcode imputation and AD imputation. In postcode imputation, we selected 20,000 records with location fields and precise postcode. Based on the postcode imputation method described in Section 4.2.1, we estimated the postcodes were estimated. The estimated postcodes were then compared with precise postcodes. The accuracy of postcode imputation was 92.18%. In AD imputation experiment, we selected 20,000 records with precise location and postcode. Based on the AD imputation method as presented in Section 4.2.2, ADs were estimated through postcodes. The estimated ADs were compared with the precise location. If the precise location contained the estimated AD information, the estimation was considered correct. The final AD imputation accuracy was 96.76%. Therefore, the imputation method obtains precise and multi-scale text addresses containing province, city, county, and street. Using this method, 97% of all records were geocoded that was much better than the method without location imputation whose geocoding rate is only 53%.

Although the location imputation achieved high accuracy and coverage rate, two kinds of deviations are still present in the final geocoding results. Geocoding sub-processes, including text address cleaning (i.e., address normalization and standardization), feature matching between inputted address and addresses in backstage reference datasets, and feature interpolation (e.g., deriving a point for an address along a street center-line or the centroid of a parcel), all may introduce deviations (Goldberg, Wilson, & Knoblock, 2007;

Roongpiboonsopit & Karimi, 2010). Another kind of deviation is derived from the differences between the registration address and operating address of an enterprise. For example, some enterprises are registered at place *A* while they are operating at place *B*. Nevertheless, both locations have significance in industrial spatial distribution analysis. Registration location reveals the regional attraction and economic activity, while the operating location has its reality in industrial spatial distribution analysis. To obtain the operating address of each enterprise, external data sources are required.

Overall, the location imputation method helps to achieve street-level accuracy, improves the geocoding rate, and supports multi-scale spatial distribution analysis of industries, including all-China-scale, inter-city-scale, city-scale, intra-city-scale, county-scale and street-scale analysis.

In general, the proposed HPC-based imputation framework is scalable, and applicable to big georeferenced text data processing that involves text classification and location imputation. Experimental results demonstrated the feasibility of Spark-based framework for high-dimensional classification problems. This work may also provide a reference for machine learning method selection in big short text classification in terms of accuracy, time and memory consumption, and scalability. Furthermore, the proposed location imputation method that integrates external spatial-associated data sources and uses NLP, is versatile and applicable to other big georeferenced text data.

### 5.2. Potential applications of imputation data

Multi-scale spatiotemporal distribution of industries in China can support the visualization and analysis of industrial aggregations, urban agglomerations, and urban spatial structures with complete, fine-grained enterprise registration data. Socioeconomic activities analysis can be extended by fusing enterprise registration data and other data sources.

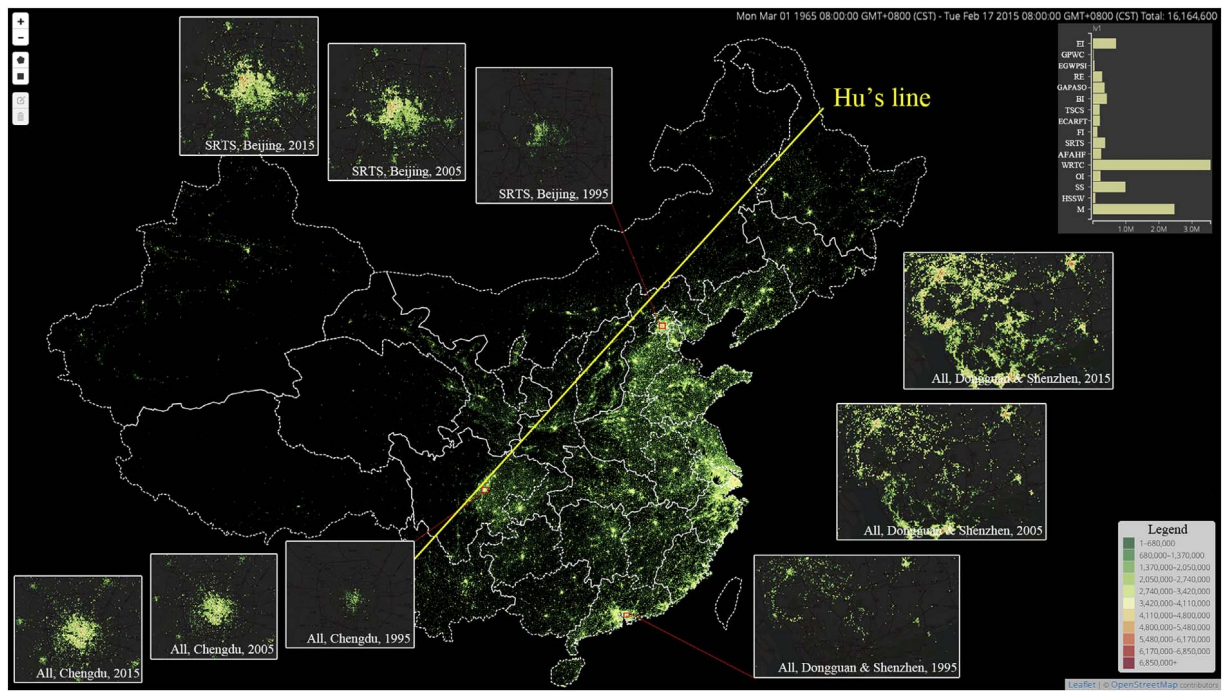

**Fig. 12.** Spatiotemporal distribution of industries in selected cities of China.

### 5.2.1. Urban changes and industrial aggregations

National-scale, inter-city-scale, and intra-city-scale industrial spatiotemporal distribution of China are briefly analyzed. In Fig. 12, the spatial distribution of all industries in China in 2015 is visualized along with the spatiotemporal distribution of industries in different cities. In China, the yellow line is Hu's (Heihe-Tengchong) line (Hu, 1935) often used in population geography. About 94.4% of population was distributed on the southeast of this line at 2010 (Chen, Gong, Li, Lu, & Zhang, 2016). About 92% of enterprises were distributed on the southeast of this line, revealing a spatial correlation between population and economic activity, as well as the sharp east-west industrial development gap in China. Three regions contain most of the enterprises as shown in Fig. 12, the Pearl River Delta region, Yangtze River Delta region, and the Beijing-Tianjin-Hebei region. Industries in Beijing, Chengdu, and Dongguan, for 1995, 2005, and 2015, are illustrated in Fig. 12. As shown, the Scientific Research and Technical Services (SRTS) in Beijing originated in the southwest of Beijing, and expanded to all of Beijing over a period of 20 years. Industries in Chengdu expanded in a radial pattern from the central city to the outskirts from 1995 to 2015. Industries in Dongguan, Guangdong province expanded from the northwest of Dongguan, close to Guangzhou, which is provincial capital of Guangdong, to southeast of Dongguan, close to Shenzhen, a special economic zone of China.

Different industries have different spatial patterns. The spatial distribution of the Chinese Agriculture, Forestry, Animal Husbandry, and Fishery (AFAHF) industry in 2013 is depicted in Fig. 13(a). The Social Service (SS) industry is shown in Fig. 13(b). The SS industry was geographical concentrated in the capital or large cities while the AFAHF industry was more evenly dispersed, revealing visually the patterns of industrial aggregations. To reveal quantitatively the geographic concentration of economic activities, we compared the aggregations of these two industries in Guangdong, China, using the widely used Ripley's K function (Giuliani, Arbia, & Espa, 2014) in economic geography. In Fig. 13(c–d), AFAHF industry is geographically concentrated within a distance of 170 km in Guangdong, while SS industry is concentrated within 520 km. SS industry is geographically concentrated within a larger space than AFAHF industry. The imputed enterprise registration data illustrates a potential use in economic geography.

### 5.2.2. Urban agglomerations and urban spatial structures

Enterprise data can support the analysis of urban agglomerations and urban spatial structure. Urban agglomerations in the three most-developed regions are illustrated in Fig. 14(a–c). These three regions are all multi-centered and each industrial center is corresponded to a city. Many enterprises connect the industrial centers. Three different urban spatial structures are illustrated in Fig. 14(d–f). Chengdu is single center structure given its flat topography. Lanzhou has a linear structure as enterprises are distributed along the river and main roads. Wuhan is a multi-centered city divided by Yangtze River and Han River. Topography, rivers, and main roads all exert influence on urban spatial structures.

### 5.2.3. Socioeconomic activities analyses by fusing multiple data source

By integrating nighttime light remote sensing data, population data, and enterprise registration data, the study of regional socioeconomic activities can be furthered. All industries, population (2010), and nighttime lights in Wuhan and Yangtze River Delta, China for 2012 are mapped in Fig. 15. VIIRS data can reveal nighttime lights and the spatial distribution of socioeconomic activities (Bennett & Smith, 2017) while population census data can depict the urban spatial structure and urban sprawl (Long & Liu, 2016; Mao, Long, & Wu, 2016; Wu, Long, Mao, & Liu, 2015). In spite of the slight deviation in the VIIRS data, census data and enterprise registration data, the overall spatial distributions of these three data sources are the same as illustrated in Fig. 15(a, b). Regional nighttime activities can be detected by the nighttime light remote sensing data; however, the composition of these activities cannot be analyzed in detail. Population census data is collected from each census statistics unit (i.e. township, town, and subdistrict) and therefore is sparser than big enterprise registration data. Individual-level big enterprise registration data plays the role of a sensor for regional socioeconomic activities and social sensing (Liu et al., 2015). By integrating these three kinds of data, regional urban studies might be rendered more precise and detailed.

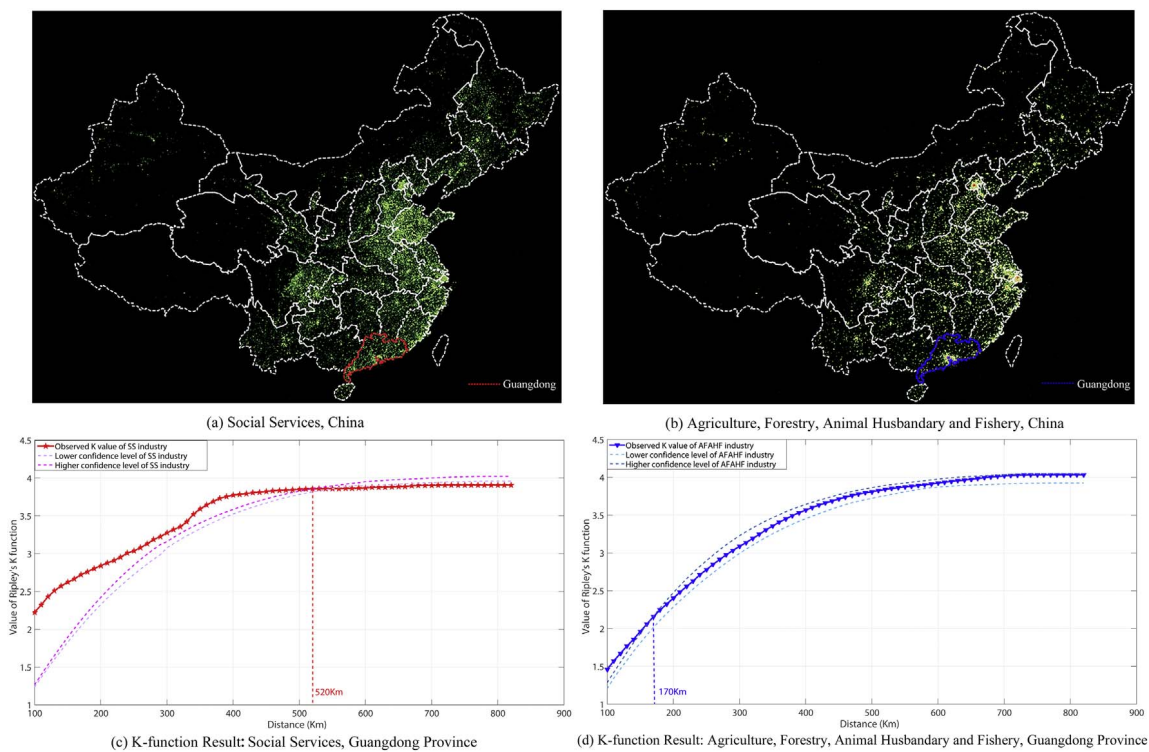

**Fig. 13.** Geographical concentration of Social Service industry and Agriculture, Forestry, Animal Husbandry, and Fishery industry in China, 2013.

Overall, complete and fine-grained enterprise registration data can enable and support multiple-scale spatiotemporal distribution analysis of Chinese industries helpful for regional development analysis and policymaking.

## 6. Conclusions and future work

We propose a big textual data imputation framework and developed several imputation methods based on the HPC technology to fill missing values in big enterprise registration data, which is also applicable for other big georeferenced text data processing. For industrial category imputation, we implemented NLP technology and compared five classification methods based on the Apache Spark framework. After exploration, we found that (1) HPC-supported imputation methods are imperative and an emerging trend in big data preprocessing. Parallel computing frameworks such as Apache Spark are scalable, and significantly reduce the execution time of big data imputation, making impossible computing tasks computable in a rational time-span or even in real-time. Parallel computing frameworks could also be helpful to overcome the Out Of Memory problems in complex computing work. (2) Big textual data missing category problems can be handled based on the proposed workflow as short textual data classification problems. When semantic information is missing from internal data sources, the integrated external data sources can provide supplementary information. (3) Machine learning methods must be carefully selected in data imputation. The Naïve Bayes method is suitable for near real-time, probability-based multiple classification problems with sparse but very high dimension input vectors; the Logistic Regression achieve higher accuracy than Naïve Bayes with longer execution time and higher memory consumption, that is more suitable for tasks with no time restrictions. For location imputation, we integrated external postcode data and implemented NLP technology to fill the missing postcodes and administrative division information. We found that such methods achieve precise text addresses including province, city, county, and street information, and largely improve geocoding proportion. The spatial distribution of generated enterprise registration data was mapped and briefly analyzed which suggests how complete and fine-grained enterprise data could well enable and support industrial spatial distribution analysis in China.

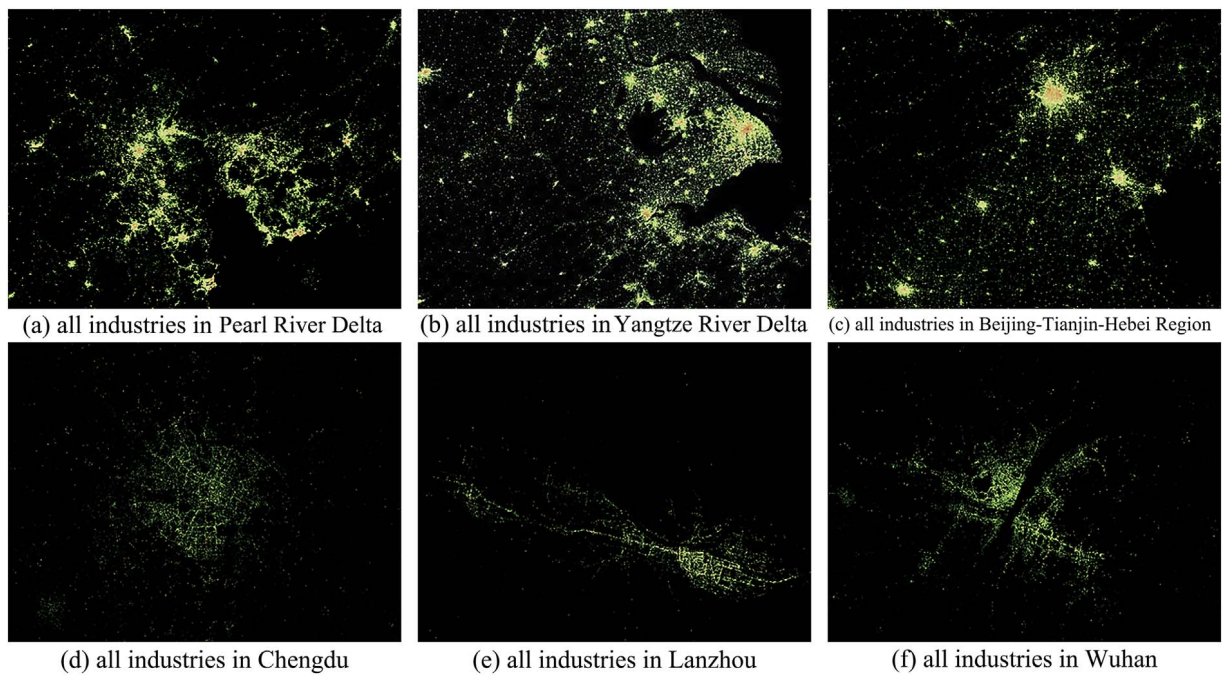

**Fig. 14.** Urban agglomerations and urban spatial structure in China, 2014.

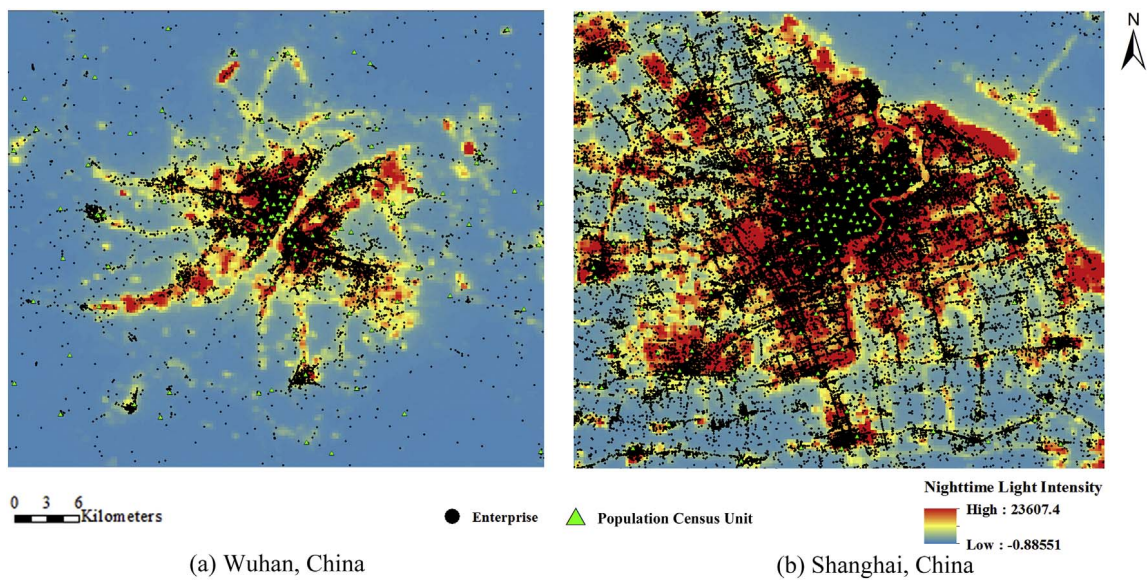

**Fig. 15.** Enterprise registration data, nighttime light data, and census data visualization.

To improve big data imputation methods and applications for developers, our potential future research direction includes:

(1) Integrating more data sources will improve industrial category imputation accuracy. Industrial category information loss in the raw data lowered the overall category imputation accuracy. By integrating data sources with industrial category related information, industrial category imputation accuracy may well be improved. For example, some websites have the complete business scope information for each enterprise. If such auxiliary information can be collected and adapted for industrial category imputation, the information loss problem could be solved and the accuracy may well be improved.

(2) Quantitative analysis of multi-scale industrial spatiotemporal distribution can support better decision making. We will quantitatively analyze Chinese industrial spatial distribution from a multi-scale, chronological, and comprehensive industrial view. To better support regional urban studies, enterprise registration data, population census data, and nighttime light data may well be integrated for a better understanding of the relationship between population, socioeconomic activities, and enterprises.

Supplementary data to this article can be found online at https://doi.org/10.1016/j.compenvurbsys.2018.01.010.

**Acknowledgements**

This paper is supported by National Key R&D Program of China (No. 2017YFB0503704 and No. 2017YFB0503802) and National Natural Science Foundation of China (No. 41501434 and No. 41371372). Thanks to Zelong Yang, Xu Gao, Xi Long, and Maoding Zhang for providing helps in system development, data preprocessing. Thanks to Stephen C. McClure for language assistance.

**Appendix A**

Table A
Correspondence between the industry categories and symbols.

| Industry category | Symbol | Industry category | Symbol |
|---|---|---|---|
| Agriculture, forestry, animal husbandry and fishery | AFAHF | Finance, insurance | FI |
| Extractive industries | EI | Real estate | RE |
| Manufacturing | M | Social services | SS |
| Electricity, gas and water production and supply industry | EGWPSI | Health, sports and social welfare | HSSW |
| Building industry | BI | Education, culture and arts, radio, film and television | ECARFT |
| Geological prospecting and water conservancy | GPWC | Scientific research and technical services | SRTS |
| Transport, storage and communications sector | TSCS | Government agencies, party agencies and social organizations | GAPASO |
| Wholesale, retail trade and catering | WRTC | Other industry | OI |

Table B

External postcode data source.

| Data source | Site link |
| --- | --- |
| Baidu postcode | http://opendata.baidu.com/post/ |
| China post | http://cpdc.chinapost.com.cn/web/ |
| Postcode library | https://www.youbianku.com/ |

Table C

Parameter setting for different classification methods.

| Classification method | Parameter | Value |
| --- | --- | --- |
| Decision Tree | Maximum depth of a tree | 30 |
| | Minimum information gain | 0.0075 |
| Random Forest | Maximum depth of a tree | 30 |
| | Minimum information gain | 0.0075 |
| | Number of trees in a forest | 100 |
| SVM | Iteration times | 50 |
| | Step size | 1 |
| | Regularization parameter | 0.01 |
| | Minimum batch fraction | 1 |

Table D

Pseudocode of Logistic Regression imputation in Apache Spark.

| | | |
| --- | --- | --- |
| 1. | conf ← SparkConf() | //configure the Spark environment |
| 2. | sc ← SparkContext(conf).textfile(train data path) | //load data from HDFS |
| 3. | fH ← HashingTF(numOfFeatures) | //initialize feature hash |
| 4. | trainData ← sc.map{ | //Map process to create training data |
| 5. | Words[1…n] ← Tokenizer (String of institution name) | //Chinese word segmentation |
| 6. | Featurewords ← null | //final feature words |
| 7. | for I from 1 to n | /*part-of-speech selection: select nouns, verb and gerund*/ |
| 8. | If words[$i$].label = n or ords[$i$].label = v or ords[$i$].label = vn | |
| 9. | Featurewords ← Featurewords + words[$i$] + " " | |
| 10. | End if | |
| 11. | End for | |
| 12. | inputVector ← LablePoint(category, fH(Featurewords))} | //construct feature vectors |
| 13. | testData ← SparkContext(conf).textfile(test data Path).map{} | //prepare test data |
| 14. | model ← LogisticRegression.run(trainData) | //train the model |
| 15. | accuracy ← model.predict(testdata) | //calculate imputation accuracy |